\documentclass{SciPost}

\hypersetup{
    colorlinks,
    linkcolor={red!50!black},
    citecolor={blue!50!black},
    urlcolor={blue!80!black}
}

\usepackage[bitstream-charter]{mathdesign}
\DeclareSymbolFont{usualmathcal}{OMS}{cmsy}{m}{n}
\DeclareSymbolFontAlphabet{\mathcal}{usualmathcal}

\fancypagestyle{SPstyle}{
\fancyhf{}
\fancyfoot[C]{\textbf{\thepage}}
}

\usepackage[T1]{fontenc}

\usepackage{orcidlink}

\usepackage{booktabs}
\usepackage{multirow}
\usepackage{tabularx}

\usepackage{listings}
\lstdefinelanguage{HLS}{
    language=C,
    morekeywords={
        ap_int,ap_fixed,
        pragma,HLS,pipeline,unroll,inline,resource,array_partition
    },
    sensitive=true
}
\usepackage[nameinlink]{cleveref}

\usepackage{csquotes}
\usepackage{textcomp}

\begin{document}

\pagestyle{SPstyle}

% ===============================================
% Title
% ===============================================
\begin{center}{\Large \textbf{\color{scipostdeepblue}{
TrackCore-F: Deploying Transformer-Based Subatomic Particle Tracking on FPGAs\\
}}}\end{center}

% ===============================================
% Authors
% ===============================================
\begin{center}\textbf{
% Authors 
% Use (full) first name (+ middle name initials) + surname format.
% Separate subsequent authors by a comma, omit comma and use "and" for the last author.
% Mark the corresponding author(s) with a superscript symbol in this order
% \star, \dagger, \ddagger, \circ, \S, \P, \parallel, ...
Arjan {Blankestijn}\textsuperscript{1}\orcidlink{0009-0005-4628-7689},
Uraz {Odyurt}\textsuperscript{2$\star$}\orcidlink{0000-0003-1094-0234} and
Amirreza {Yousefzadeh}\textsuperscript{1}\orcidlink{0000-0002-2967-5090}
}\end{center}

\begin{center}
% Affiliations
{\bf 1} Computer Architecture for Embedded Systems, University of Twente, Enschede, The Netherlands
\\
{\bf 2} Faculty of Engineering Technology, University of Twente, Enschede, The Netherlands
% Email address of corresponding author(s)
\\[\baselineskip]
$\star$ \href{mailto:u.odyurt@utwente.nl}{\small u.odyurt@utwente.nl}\,
%,\quad
%$\dagger$ \href{mailto:email2}{\small email2}
\end{center}

% ===============================================
% Conference details
% ===============================================
\definecolor{palegray}{gray}{0.95}
\begin{center}
\colorbox{palegray}{
  \begin{tabular}{rr}
  \begin{minipage}{0.37\textwidth}
    \includegraphics[width=60mm]{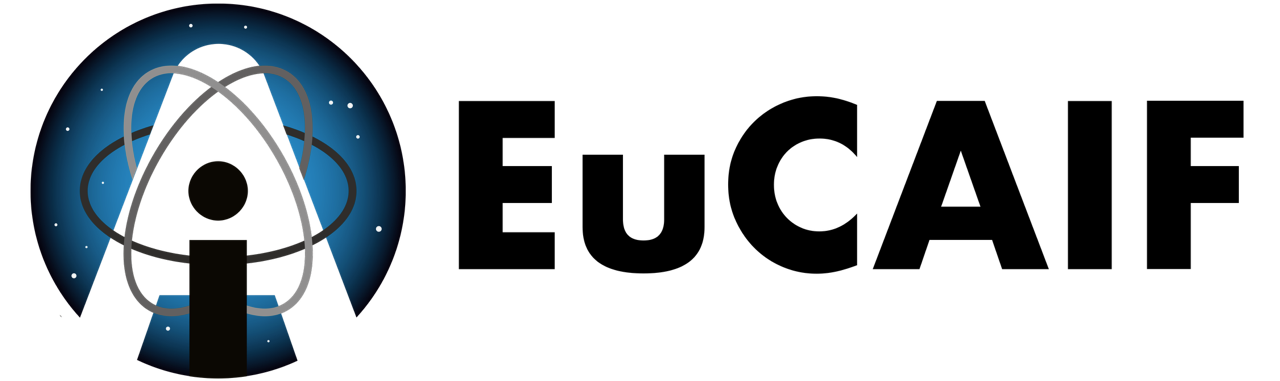}
  \end{minipage}
  &
  \begin{minipage}{0.5\textwidth}
    \vspace{5pt}
    \vspace{0.5\baselineskip} 
    \begin{center} \hspace{5pt}
    {\it The 2nd European AI for Fundamental \\Physics Conference (EuCAIFCon2025)} \\
    {\it Cagliari, Sardinia, 16-20 June 2025
    }
    \vspace{0.5\baselineskip} 
    \vspace{5pt}
    \end{center}
    
  \end{minipage}
\end{tabular}
}
\end{center}

% ###############################################
% Abstract
% ###############################################
\section*{\color{scipostdeepblue}{Abstract}}
\textbf{\boldmath{%
The Transformer Machine Learning (ML) architecture has been gaining considerable momentum in recent years. In particular, computational High-Energy Physics tasks such as jet tagging and particle track reconstruction (tracking), have either achieved proper solutions, or reached considerable milestones using Transformers. On the other hand, the use of specialised hardware accelerators, especially FPGAs, is an effective method to achieve online, or pseudo-online latencies. The development and integration of Transformer-based ML to FPGAs is still ongoing and the support from current tools is very limited or non-existent. Additionally, FPGA resources present a significant constraint. Considering the model size alone, while smaller models can be deployed directly, larger models are to be partitioned in a meaningful and ideally, automated way. We aim to develop methodologies and tools for monolithic, or partitioned Transformer synthesis, specifically targeting inference. Our primary use-case involves two machine learning model designs for tracking, derived from the TrackFormers project. We elaborate our development approach, present preliminary results, and provide comparisons.
}}

\vspace{\baselineskip}

% ===============================================
% Copyright
% ===============================================
% This block will be filled during the proof stage, and finilized just before publication.
% It exists here only as a placeholder, and should not be modified by authors.
%\noindent\textcolor{white!90!black}{%
%\fbox{\parbox{0.975\linewidth}{%
%\textcolor{white!40!black}{\begin{tabular}{lr}%
%  \begin{minipage}{0.6\textwidth}%
%    {\small Copyright attribution to authors. \newline
%    This work is a submission to SciPost Phys. Proc. \newline
%    License information to appear upon publication. \newline
%    Publication information to appear upon publication.}
%  \end{minipage} & \begin{minipage}{0.4\textwidth}
%    {\small Received Date \newline Accepted Date \newline Published Date}%
%  \end{minipage}
%\end{tabular}}
%}}
%}

% ===============================================
% Line numbers
% ===============================================
%\linenumbers

% ===============================================
% Table of contents
%\vspace{10pt}
%\noindent\rule{\textwidth}{1pt}
%\tableofcontents
%\noindent\rule{\textwidth}{1pt}
%\vspace{10pt}

% ###############################################
% Text body
% ###############################################
% ###############################################
% Start of file - body.tex
% ###############################################

% ===============================================
% Section
% ===============================================
\section{Introduction}
\label{sec:introduction}
The move to ML-assisted tracking solutions is deemed necessary and inevitable. While design efforts are moving forward, another important aspect to consider is deployment. The accurate tracking has been a post-mortem task, which is due to the computational requirements. While the introduction of ML algorithms will improve the computational performance, the increase in scale and frequency of experiments will somewhat counter and reduce effects of such improvements. This is especially the case for the scales expected from the upcoming High-Luminosity stage of the LHC (HL-LHC). On top of that, efficient execution of ML algorithms has predominantly been tied to Graphics Processing Units (GPUs) as the platform of choice. While not a universal dependency, GPUs dominate the ML deployment scene. Nevertheless, there are viable alternative forms of hardware acceleration, e.g., Field-Programmable Gate Array (FPGA), custom Application-Specific Integrated Circuit (ASIC), and Neuromorphic. We aim to deploy ML-assisted track reconstruction on FPGAs to achieve better or equivalent latencies. Additionally, FPGAs enable on-site deployment of tracking and entail considerable energy efficiency potential.

\paragraph*{Related work}
Deployment of models based on the Transformer architecture on hardware, especially FPGAs, has been a point of interest for the research community. This is especially noticeable during the past and the current year, 2024--2025. Focusing on the optimisations, 4 main trends are observable, namely: approaches focusing on quantization techniques~\cite{Guo:2025:VTAH, Zhao:2025:EVTA, Du:2024:EEFB}, approaches utilising off-chip memory~\cite{Kabir:2025:RATN}, approaches leveraging sparsity through pruning~\cite{Zhang:2024:EFTA, Wang:2024:EFTA, Li:2024:EEFA}, and approaches based on optimisation of non-linear functions such as SoftMax~\cite{Bai:2024:SEWA}.

One of the important metrics to consider when evaluating implementations is \emph{throughput}. At the time of this writing, the most promising approaches demonstrating the highest throughput are~\cite{Guo:2025:VTAH} and~\cite{He:2025:FTFA}, with the former achieving noticeably higher throughput.

% ===============================================
% Section
% ===============================================
\section{Background}
\label{sec:background}
At the Large Hadron Collider (LHC), particles are accelerated in opposite directions in a circular accelerator and made to collide at four interaction points, where large-scale detectors are positioned. These major detectors are ALICE~\cite{Collaboration:2008:ALICE}, ATLAS~\cite{Collaboration:2008:ATLAS}, CMS~\cite{Collaboration:2008:CMS}, and LHCb~\cite{Collaboration:2008:LHCb}. Detectors perform two key functions: \emph{tracking} and \emph{calorimetry}, allowing the calculation of particle momentum, $p$ and measuring the energy, $E$, deposited by particles, respectively. Together, these measurements enable the calculation of a particle's mass, $m$, using the relativistic energy-momentum relation: $E^2 = (mc^2)^2 + (pc)^2$, where $c$ is the speed of light. Accurately determining particle mass is essential for identifying known particles and discovering new ones.

\subsection{Tracking algorithms}
There has been continuous efforts channelled into the design and development of ML-based, or rather ML-assisted, tracking solutions. Two ML model architectures stand out: Graph Neural Networks (GNNs) and more recently, Transformers~\cite{Vaswani:2017:Attention}. Within the scope of this paper, we focus on two Transformer designs from the project TrackFormers~\cite{Caron:2025:TrackFormers}, \emph{EncCla} and \emph{EncReg}. Both models operate as so-called single-shot models, i.e., they take in a full event's data and perform hit to track association for the whole event.

The Encoder-Classifier (EncCla) is an encoder-only Transformer design. This approach takes in hit coordinates from an event and predicts their association to pre-defined class labels. Class labels, ensured to be unique, are generated through binning of the track parameter space. The largest variant has close to 1.5 million parameters with estimated memory consumptions of 5.69 MB and 0.07 MB for parameters and activations, respectively.

The Encoder-Regressor (EncReg) is also an encoder-only Transformer design. It does not rely on class labels, but instead regresses the parameters of potential tracks for a single event. As a post-processing step, a clustering algorithm has to be applied to model's output. As such, predicted track parameters per hit are clustered, forming track associations. EncReg uses HDBSCAN to achieve this clustering. The model has close to 76\,484 parameters with estimated memory consumptions of 0.29 MB and 0.07 MB for parameters and activations, respectively.

\subsection{Datasets}
\label{subsec:datasets}
The two model designs have been trained with 5 different datasets, forming a progression of simple to complex representations of tracks and hits, i.e., track function complexity, track count, and by extension, hit count: 10--50 (variable count) linear tracks per event (REDVID), 10--50 (variable count) helical tracks per event (REDVID), 50--100 (variable count) helical tracks per event (REDVID), 10--50 (variable count) tracks per event (TrackML), 200--500 (variable count) tracks per event (TrackML). The first three datasets are the result of simulations using REDVID simulation framework~\cite{Odyurt:2024:RSHE}. The last two datasets are scale-reduced versions of the data associated with the TrackML Kaggle challenge~\cite{Kiehn:2019:TrackML}.

% ===============================================
% Section
% ===============================================
\section{Implementation and results}
\label{sec:implementation}
The utilised test bench is an ARM Zynq UltraScale+ MPSoC ZCU102 evaluation kit. The on-board EG device (ZU9EG) consists of a Quad-Core ARM Cortex-A53 Processing Unit (PU) and a Programmable Logic (PL) with the following specification~\cite{AMD:2025:Zynq}: System Logic Cells 599\,550, Configurable Logic Block (CLB) Flip-Flops 548\,160, CLB LUTs 274\,080, Distributed RAM (Mb) 8.8, Block RAM (Mb) 32.1, DSP Slices 2\,520.

A variety of tools must be used in sequence to obtain a deployable synthesised kernel. In particular, for pre-trained ML models, the following have been used:
\begin{itemize}
    \item PyTorch: Models are provided in the PyTorch format, which could optionally be used for quantization.
    \item ONNX: Open Neural Network Exchange (ONNX) is an open-source format for representing ML models, which reveals low-level operations, input/output dimensions, data types, and weight Tensor values (if present) per operation. The ONNX format is utilised for quantization as well.
    \item AMD Vitis HLS 2022.2: Vitis High-Level Synthesis (HLS) is a tool from AMD (previously Xilinx), enabling C/C++, or OpenCL descriptions of hardware to be synthesised into Register Transfer Level (RTL) implementations targeting FPGAs.
    \item AMD Vivado 2022.2: Vivado Design Suite is used to integrate and configure the IP designed using Vitis HLS with the Zynq MPSoC and to synthesise the final bitstream. 
    \item PYNQ: Python Productivity for Zynq (PYNQ) is a Python-based development environment designed for AMD's Zynq SoCs. PYNQ enables interaction with the Zynq PL.
\end{itemize}

\subsection{Development flow}
\begin{figure}[htbp]
	\centering
	\includegraphics[width=\linewidth]{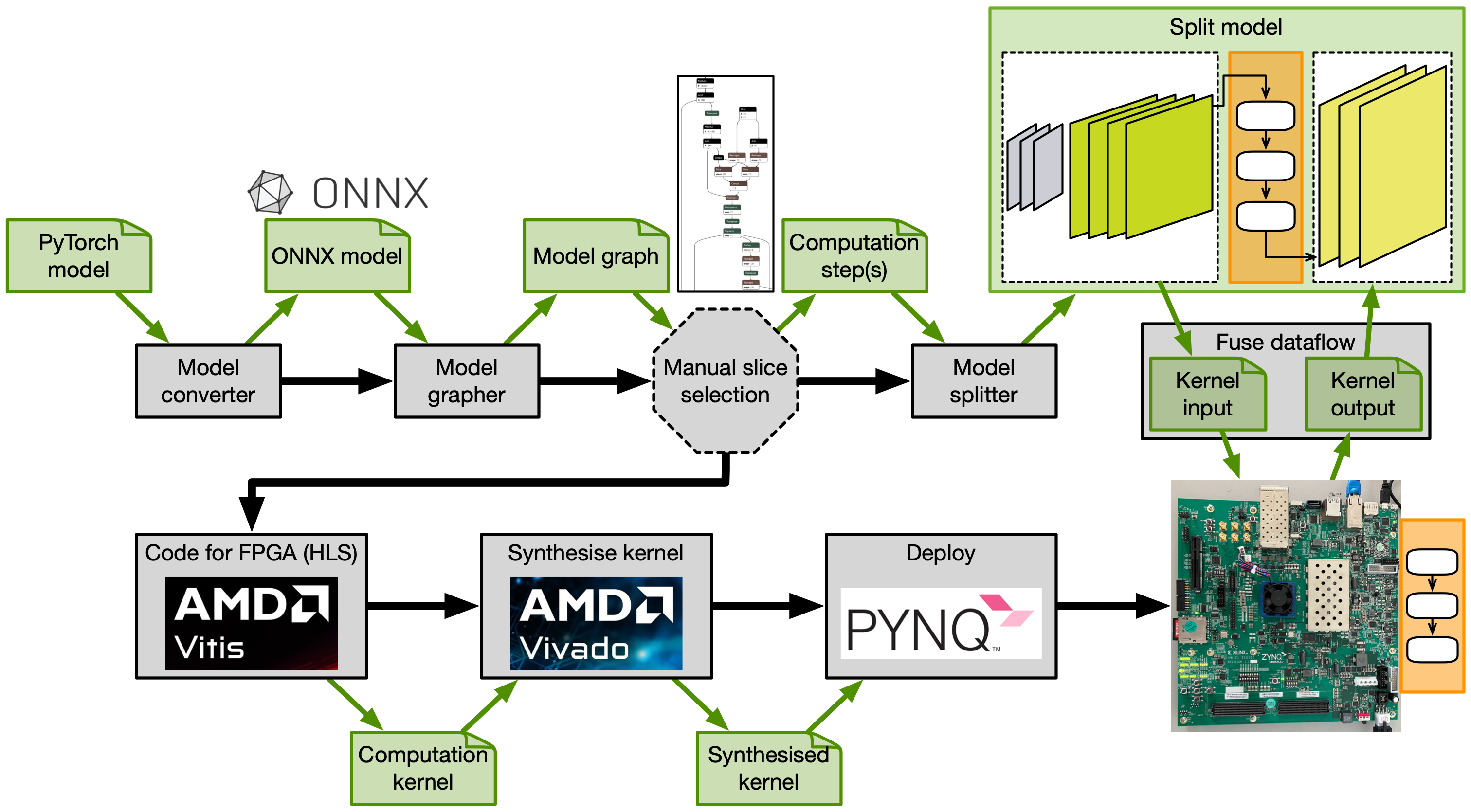}
	\caption{The development flow describing the handling of pre-trained ML models and preparations for selective slice deployment on a FPGA.}
	\label{fig:development_flow}
\end{figure}

As depicted in \Cref{fig:development_flow}, our deployment workflow is comprised of different steps involving the aforementioned tools.

First, a pre-trained PyTorch model is converted to the ONNX format. The selected model is trained on the first dataset from \Cref{subsec:datasets}. Working with an ONNX model requires the model itself, plus the input data shape. Considering the model graph of low-level computational operations (through ONNX format), one can opt for a full or a partial model deployment. A full deployment is conditional to model size and PL resource availability.

Currently, we apply a manual partitioning, which results in a split model, including the slice to be synthesised as a kernel. The output from the preceding layer is passed on to the slice kernel. In return, the output from the kernel is being fed to the following layer, fusing the dataflow between model partitions. Vitis HLS is used to develop the kernel in C/C++. Once functionality is verified with behavioural simulation, the kernel is synthesised into a hardware IP. The resulting IP is imported into Vivado where it is integrated with the MPSoC block and connected with peripherals such as the AXI4 communication bus. Finally, using Vivado, the complete design is synthesised and a bitstream generated. Note that our current workflow relies on the Vivado IP Flow to integrate our designed IP into the rest of the peripherals using Vivado. Optionally, the (more restrictive) alternative Vitis Kernel Flow can be used to directly synthesise the final output.

Focusing on a single encoder layer, the model is split into three parts: the segment before the first encoder layer, the encoder layer to be implemented and deployed on the FPGA, and the remaining model segment. The encoder kernel block is developed using HLS C/C++ in Vitis. The top-level function takes all the weights of the encoder layer as input. These weights are transferred to the kernel using an AXI4-Lite interface. Once all the parameters are loaded in, the computations for the encoder layer take place. This consists of the Multi-Head Attention (MHA), followed by the first \texttt{Add AND Normalisation}, a feed-forward layer, and finally a last \texttt{Add AND Normalisation} layer. HLS Pragmas are used to optimise performance.

When designing a kernel for the PL, we are not required to implement ONNX operation blocks individually. In fact, it often reduces CLB and memory utilisation if a selected segment can be developed into a monolithic kernel. To make the transition and porting smooth, we first develop the ONNX model slice using low-level Python code, i.e., NumPy and basic arithmetic.

\subsection{Quantization effects}
\label{subsec:quantization_effects}
The current implementation makes use of floating-point weights and activations. A common way to increase performance and reduce resource consumption is to employ quantization, often at \texttt{INT8}, for both weights and activations. However, this can greatly affect the overall accuracy of the model. ONNX Static quantization available in the ONNX Python package was used to investigate the effect on the accuracy. \Cref{tab:quantizations} shows the resulting model accuracies at different quantization configurations. We can conclude that quantizing the activations has a bigger impact on the model accuracy, while quantizing just the weights has a lesser impact. 
\begin{table}[htbp]
    \centering
    \caption{Model prediction accuracies with different quantization levels applied. The base model has an overall prediction accuracy of 0.97.}
    \label{tab:quantizations}
    \begin{tabularx}{0.7\linewidth}{@{}llc@{}}
        \toprule
        \textbf{Model activations} & 
        \textbf{Model weights} & 
        \textbf{Prediction accuracy} \\
        \midrule
        \texttt{INT16}   & \texttt{INT16}     & 0.90 \\
        \texttt{INT16}   & \texttt{INT8}      & 0.90 \\
        \texttt{INT8}    & \texttt{INT16}     & 0.71 \\
        \texttt{INT8}    & \texttt{INT8}      & 0.70 \\
        \bottomrule
    \end{tabularx}
\end{table}

There are more combinations to try for a comprehensive benchmarking, e.g., not quantizing the activations and only using quantized \texttt{INT8} weights. ONNX Static quantization does not support this configuration and implementation is more complicated.

\subsection{Resource consumption}
The FPGA has a limited amount of available hardware resources, in particular, Block RAMs (BRAMs), Digital Signal Processing slices (DSPs), Flip-Flops (FFs), and Look-Up Tables (LUTs). \Cref{tab:resource-utilisation} lists the resource consumption of a single encoder layer kernel compared to the available resources on the ZCU102.
\begin{table}[htbp]
    \centering
    \caption{ZCU102 FPGA resource utilisation for a single encoder layer, as reported by Vitis.}
    \label{tab:resource-utilisation}
    \begin{tabularx}{0.7\linewidth}{@{}lccc@{}}
        \toprule
        \textbf{Resource type} & 
        \textbf{Used} & 
        \textbf{Total available} & 
        \textbf{Utilisation (\%)} \\
        \midrule
        36 Kb BRAM		& 347 					& 912 			& 38.04 \\
        DSP     		& 67 					& 2\,520      	& 2.66 \\
        FF      		& 74\,184   			& 548\,160    	& 13.53 \\
        LUT     		& 65\,815    			& 274\,080    	& 24.01 \\
% ---------------------------
%36 Kb BRAM		& 347 (344.50) 			& 912 			& 38.04 (37.77) \\
%DSP     		& 67 (76)				& 2\,520      	& 2.66 (3.01) \\
%FF      		& 74\,184 (57\,039)  	& 548\,160    	& 13.53 (10.40) \\
%LUT     		& 65\,815 (39\,487)   	& 274\,080    	& 24.01 (14.41) \\
% ---------------------------
        \bottomrule
    \end{tabularx}
\end{table}

As it can be seen, the limiting factor is the available BRAM, which has a utilisation of 38.04\%. However, the number of used BRAMs can be reduced effectively by considering more DDR memory at the expense of increased memory transfer and latency. Thus, given that the BRAM utilisation can be reduced, the second limiting factor will be the number of LUTs, which is at 24.01\% utilisation for a single encoder layer. This suggests that a maximum of 4 encoder layers can be implemented on this particular FPGA and for this model.

% ===============================================
% Section
% ===============================================
\section{Conclusion}
\label{sec:conclusion}
We have presented a structured development flow for the deployment of pre-trained tracking models on FPGAs. A partial deployment is as feasible as a full deployment. Although not as performant, we consider a partial deployment as valuable as a full one, since it enables model inference to be deployed on more accessible hardware. We observe that the application of optimisations can be more important than the HLS implementation itself. We have also shown how detrimental quantization can be to the model's prediction accuracy. Optimisations and their effects are highly dependent on the model and the HLS implementation.

% ===============================================
% Acknowledgements
% ===============================================
\section*{Acknowledgements}
This publication is part of the project ZORRO with project number KICH1.ST02.21.003 of the research programme Key Enabling Technologies (KIC), which is (partly) financed by the Dutch Research Council (NWO).

% ###############################################
% End of file
% ###############################################

% ###############################################
% Bibliography
% ###############################################
\bibliography{bibliography/references.bib}

@article{Collaboration:2008:ALICE,
	author = {The ALICE Collaboration},
	title = {{The ALICE experiment at the CERN LHC}},
	journal = {Journal of Instrumentation},
	year = {2008},
	OPTmonth = Aug,
	OPTpublisher = {},
	OPTvolume = {3},
	OPTnumber = {08},
	OPTpages = {S08002},
	doi = {10.1088/1748-0221/3/08/S08002}
}

@article{Collaboration:2008:CMS,
	author = {The CMS Collaboration},
	title = {{The CMS experiment at the CERN LHC}},
	journal = {Journal of Instrumentation},
	year = {2008},
	OPTmonth = {aug},
	OPTpublisher = {},
	OPTvolume = {3},
	OPTnumber = {08},
	OPTpages = {S08004},
	doi = {10.1088/1748-0221/3/08/S08004}
}

@article{Collaboration:2008:LHCb,
	author = {The LHCb Collaboration},
	title = {{The LHCb Detector at the LHC}},
	journal = {Journal of Instrumentation},
	year = {2008},
	OPTmonth = {aug},
	OPTpublisher = {},
	OPTvolume = {3},
	OPTnumber = {08},
	OPTpages = {S08005},
	doi = {10.1088/1748-0221/3/08/S08005}
}

@article{Collaboration:2008:ATLAS,
	author = {The ATLAS Collaboration},
	title = {{The ATLAS Experiment at the CERN Large Hadron Collider}},
	journal = {Journal of Instrumentation},
	year = {2008},
	OPTmonth = {aug},
	OPTpublisher = {},
	OPTvolume = {3},
	OPTnumber = {08},
	OPTpages = {S08003},
	doi = {10.1088/1748-0221/3/08/S08003}
}

@inproceedings{Vaswani:2017:Attention,
    author = {Vaswani, Ashish and Shazeer, Noam and Parmar, Niki and Uszkoreit, Jakob and Jones, Llion and Gomez, Aidan N. and Kaiser, \L{}ukasz and Polosukhin, Illia},
    title = {{Attention is All You Need}},
    year = {2017},
    OPTisbn = {9781510860964},
    OPTpublisher = {Curran Associates Inc.},
    OPTaddress = {Red Hook, NY, USA},
    booktitle = {Proceedings of the 31st International Conference on Neural Information Processing Systems},
    OPTpages = {6000–6010},
    OPTnumpages = {11},
    OPTlocation = {Long Beach, California, USA},
    OPTseries = {NIPS'17}
}

@article{Caron:2025:TrackFormers,
    author = {Sascha {Caron} and Nadezhda {Dobreva} and Antonio {Ferrer Sánchez} and José D. {Martín-Guerrero} and Uraz {Odyurt} and Roberto {Ruiz de Austri Bazan} and Zef {Wolffs} and Yue {Zhao}},
    title = {{TrackFormers: In Search of Transformer-Based Particle Tracking for the High-Luminosity LHC Era}},
    journal = {The European Physical Journal C},
    OPTvolume = {85},
    OPTnumber = {4},
    OPTpages = {460},
    year = {2025},
    OPTdate = {2025-04-25},
    OPTissn = {1434-6052},
    doi = {10.1140/epjc/s10052-025-14156-3}
}

@inproceedings{Odyurt:2024:RSHE,
	author = {Odyurt, Uraz and Swatman, Stephen Nicholas and Varbanescu, Ana-Lucia and Caron, Sascha},
	OPTeditor = {Franco, Leonardo and de Mulatier, Cl{\'e}lia and Paszynski, Maciej and Krzhizhanovskaya, Valeria V. and Dongarra, Jack J. and Sloot, Peter M. A.},
	title = {{Reduced Simulations for High-Energy Physics, a Middle Ground for Data-Driven Physics Research}},
	booktitle = {Computational Science -- ICCS 2024},
	year = {2024},
	OPTpublisher = {Springer Nature Switzerland},
	OPTaddress = {Cham},
	OPTpages = {84--99},
	OPTisbn = {978-3-031-63751-3},
	doi = {10.1007/978-3-031-63751-3_6}
}

@article{Kiehn:2019:TrackML,
	author = {{Kiehn, Moritz} and {Amrouche, Sabrina} and {Calafiura, Paolo} and {Estrade, Victor} and {Farrell, Steven} and et al.},
	title = {{The TrackML high-energy physics tracking challenge on Kaggle}},
	journal = {EPJ Web Conf.},
	year = {2019},
	OPTvolume = {214},
	OPTpages = {06037},
    doi = {10.1051/epjconf/201921406037}
}

@manual{AMD:2025:Zynq,
    title = {{Zynq UltraScale+\ MPSoC Data Sheet: Overview}},
    author = {{AMD Technical Information Portal}},
    organization = {{AMD}},
    type = {Technical Data Sheet},
    number = {DS891 (v1.11.1)},
    date = {2025-03-18},
    url = {https://docs.amd.com/v/u/en-US/ds891-zynq-ultrascale-plus-overview}
}

@inproceedings{Guo:2025:VTAH,
    author = {Guo, Qingyu and Wan, Jiayong and Xu, Songqiang and Li, Meng and Wang, Yuan},
    title = {{HG-PIPE: Vision Transformer Acceleration with Hybrid-Grained Pipeline}},
    booktitle = {Proceedings of the 43rd IEEE/ACM International Conference on Computer-Aided Design},
    OPTarticleno = {120},
    OPTnumpages = {9},
    OPTlocation = {Newark Liberty International Airport Marriott, New York, NY, USA},
    OPTseries = {ICCAD '24},
    year = {2025},
    OPTisbn = {9798400710773},
    OPTpublisher = {Association for Computing Machinery},
    OPTaddress = {New York, NY, USA},
    doi = {10.1145/3676536.3676681}
}

@misc{Kabir:2025:RATN,
    title = {{A Runtime-Adaptive Transformer Neural Network Accelerator on FPGAs}}, 
    author = {Ehsan Kabir and Jason D. Bakos and David Andrews and Miaoqing Huang},
    year = {2025},
    OPTeprint = {2411.18148},
    OPTarchivePrefix = {arXiv},
    OPTprimaryClass = {cs.AR},
    doi = {10.48550/arXiv.2411.18148},
    OPTurl = {https://arxiv.org/abs/2411.18148}
}

@article{Zhao:2025:EVTA,
    author = {Zhao, Pan and Xue, Donghui and Wu, Licheng and Chang, Liang and Tan, Haining and Han, Yinhe and Zhou, Jun},
    title = {{HEAT: Efficient Vision Transformer Accelerator With Hybrid-Precision Quantization}}, 
    journal = {IEEE Transactions on Circuits and Systems II: Express Briefs}, 
    year = {2025},
    OPTvolume = {72},
    OPTnumber = {5},
    OPTpages = {758-762},
    doi = {10.1109/TCSII.2025.3547340}
}

@inproceedings{Du:2024:EEFB,
    author = {Du, Congpeng and Ko, Seok-Bum and Zhang, Hao},
    title = {{Energy Efficient FPGA-Based Binary Transformer Accelerator for Edge Devices}}, 
    booktitle = {2024 IEEE International Symposium on Circuits and Systems (ISCAS)}, 
    year = {2024},
    OPTvolume = {},
    OPTnumber = {},
    OPTpages = {1-5},
    doi = {10.1109/ISCAS58744.2024.10558631}
}

@inproceedings{Zhang:2024:EFTA,
    author = {Zhang, Manting and Cao, Jialin and Shi, Kejia and Zhao, Keqing and Zhang, Genhao and Yu, Jun and Wang, Kun},
    title = {{FNM-Trans: Efficient FPGA-based Transformer Architecture with Full N:M Sparsity}},
    booktitle = {Proceedings of the 61st ACM/IEEE Design Automation Conference},
    OPTarticleno = {191},
    OPTnumpages = {6},
    OPTlocation = {San Francisco, CA, USA},
    OPTseries = {DAC '24},
    year = {2024},
    OPTisbn = {9798400706011},
    OPTpublisher = {Association for Computing Machinery},
    OPTaddress = {New York, NY, USA},
    doi = {10.1145/3649329.3656497}
}

@inproceedings{Wang:2024:EFTA,
    author = {Wang, Saiqun and Zhang, Hao}, 
    title = {{Efficient FPGA-Based Transformer Accelerator Using In-Block Balanced Pruning}}, 
    booktitle = {2024 13th International Conference on Communications, Circuits and Systems (ICCCAS)},
    year = {2024},
    OPTvolume = {},
    OPTnumber = {},
    OPTpages = {18-23},
    doi = {10.1109/ICCCAS62034.2024.10651591}
}

@inproceedings{Li:2024:EEFA,
    author = {Li, Zuohao and Lai, Yiwan and Zhang, Hao},
    title = {{Energy Efficient FPGA-Based Accelerator for Dynamic Sparse Transformer}}, 
    booktitle = {2024 13th International Conference on Communications, Circuits and Systems (ICCCAS)}, 
    year = {2024},
    OPTvolume = {},
    OPTnumber = {},
    OPTpages = {7-12},
    doi = {10.1109/ICCCAS62034.2024.10652850}
}

@inproceedings{Bai:2024:SEWA,
    author = {Bai, Zhenyu and Dangi, Pranav and Li, Huize and Mitra, Tulika},
    title = {{SWAT: Scalable and Efficient Window Attention-based Transformers Acceleration on FPGAs}},
    booktitle = {Proceedings of the 61st ACM/IEEE Design Automation Conference},
    OPTarticleno = {93},
    OPTnumpages = {6},
    OPTlocation = {San Francisco, CA, USA},
    OPTseries = {DAC '24},
    year = {2024},
    OPTisbn = {9798400706011},
    OPTpublisher = {Association for Computing Machinery},
    OPTaddress = {New York, NY, USA},
    doi = {10.1145/3649329.3658488}
}

@article{He:2025:FTFA,
    author={He, Zerong and Jin, Xi and Xu, Zhongguang},
    title={{F3: An FPGA-Based Transformer Fine-Tuning Accelerator With Flexible Floating Point Format}}, 
    journal={IEEE Journal on Emerging and Selected Topics in Circuits and Systems}, 
    year={2025},
    OPTvolume={15},
    OPTnumber={2},
    OPTpages={258-271},
    doi={10.1109/JETCAS.2025.3555970}
}

% ###############################################
% Document end
% ###############################################
\end{document}